\documentclass[smallextended]{svjour3}       
\smartqed  
\usepackage{graphicx}
\usepackage{gensymb}
\usepackage{color}
\usepackage{siunitx}

\begin{document}

\title{Perspectives of a visible instrument on the VLTI\thanks{This paper has benefited from European funding through the H2020 OPTICON program, with the grant agreement n$^\circ$730 890}}

\author{Florentin Millour         \and
        Denis Mourard \and
        Julien Woillez \and
        Philippe Berio \and
        Andrea~Chiavassa \and
        Orlagh Creevey \and
        Eric~Lagadec \and
        Marc-Antoine Martinod \and
        Anthony Meilland \and
        Nicolas Nardetto \and
        Karine Perraut \and
        Philippe Stee
        }

\authorrunning{F. Millour et al.}

\institute{F. Millour, D. Mourard, P. Berio, A. Chiavassa, O. Creevey, E. Lagadec, M. A. Martinod, A. Meilland, N. Nardetto, and P. Stee work all at \at
              Universit\'{e} C\^{o}te d'Azur, OCA, CNRS, Lagrange, Parc Valrose, B\^{a}t. Fizeau, 06108 Nice cedex 02, France
              \email{fmillour@oca.eu}
           \and
           J. Woillez \at
  ESO, Karl-Schwarzschild-Strasse 2,
D-85748 Garching bei M\"unchen
           \and
           K. Perraut \at
  Universit\'{e} Grenoble Alpes, CNRS, IPAG, 38000 Grenoble, France
}

\date{Received: date / Accepted: date}

\maketitle

\begin{abstract}
In this paper we present the most promising science cases for a new generation visible instrument on the VLTI and the conceptual idea for the instrumental configuration. We also present a statistical study of the potential targets that may be accessible for the different classes of objects and for the required spectral resolutions.
\keywords{high angular resolution \and visible interferometry \and fundamental stellar parameters \and stellar imaging}
\end{abstract}


\section{Introduction}
\label{intro}

The last decade of Optical Interferometry has seen an important expansion of the astrophysical results coming mainly from the VLTI \cite{vlti} and the CHARA arrays \cite{chara}. At the same time, the expansion of the user's community has progressed with the preparation of the second generation instruments \mbox{GRAVITY} \cite{gravity2017} and MATISSE \cite{matisse} on one side, and the opening of CHARA to the wider community on the other side. Both arrays will be equipped soon with adaptive optics systems on all their telescopes.

While science programs have continued to be well diversified, it is also clear that significant efforts have been devoted to the determination of fundamental stellar parameters, with an important complementarity with asteroseismology, spectroscopy, photometry, and now astrometry \cite{gaia}. In a recent work \cite{stee2017}, Stee et al. produced a review of the potential science cases for optical interferometry at visible wavelengths. With the launch of new space missions dedicated to stars and planets (TESS, CHEOPS, PLATO), optical interferometry has an important role to play in bringing direct measurements of stellar parameters. They will permit one to correctly disentangle the planet parameters from the potential stellar activity. It is clear however that not only the best possible angular resolution will be necessary to correctly sample the selected stars, but also good imaging quality will be required to correctly characterize the stellar surface features. The long baselines of the CHARA Array and the dense configurations of the VLTI are thus very complementary.

In Mourard et al. (2017) \cite{spica2017}, we presented our perspective for a new 6-telescope beam combiner for the CHARA Array. In the present paper, we describe our ideas to develop a visible instrument for the VLTI: Sec.~\ref{sec:sciencecases} presents the most promising science cases, Sec.~\ref{sec:instru} details the main instrumental requirements that will be necessary to achieve our science goals, and Sec.~\ref{sec:targets} presents a study of the potential samples of targets in the different classes.

\section{Unique science that can be done at VLTI in the visible}
\label{sec:sciencecases}

In 2012, O. Chesneau started a prospective work in the \emph{Action Sp\'ecifique Haute R\'esolution Angulaire} (ASHRA) from the \emph{Centre National de la Recherche Scientifique} (CNRS). The initial idea was to make a past, present and future appraisal of the type of science that can be addressed with visible interferometry. This work culminated with the publication of the visible interferometry white book \cite{stee2017}.

High spatial resolution is the key to understanding various astrophysical phenomena. But even with the future ELT, single dish instruments are limited to a spatial resolution of about 4 mas in the visible. As a reference point, most stellar photospheres in our Galaxy remain smaller than 1 mas.


We started exploiting this whitebook with the SPICA project on the CHARA array. 
SPICA will provide many improvements compared to now: 6 telescopes, full use of the new adaptive optics, use optical fibers to perform spatial filtering, fringe tracking, and the use of new low noise EMCCD cameras. It will focus on two main science cases: an angular diameter survey of all observable stars up to the magnitude $m_R=10$, and an imaging survey program to produce surfaces images in the visible of giant and supergiant stars. This instrument will benefit from two unique features of CHARA : the longest operating baselines in the world (330\,m), and 6 telescopes at full throttle. It is also considered as a testbench for the technologies of a future visible VLTI visible instrument.

Indeed, the VLTI has very complementary characteristics compared to CHARA, which makes it very appealing for hosting a visible instrument: the telescopes are larger (1.8\,m and 8\,m), there are shorter baselines (starting at 8\,m), allowing a better sampling of small spatial frequencies, and the array is denser (due to the movable auxiliary telescopes). The Paranal site has also an excellent median seeing of 0.8" in the visible domain, and clear skies all over the year. In addition, the arrival of the second-generation instruments of the VLTI triggered a thorough assessment and improvement of the VLTI infrastructure performance \cite{Woillez+2016,Woillez+2018}, making relevant the perspective of a visible addition to the VLTI instruments suite.

In summary, while CHARA reaches a higher angular resolution, the VLTI has an advantage for imaging larger targets (AT and short baselines), and fainter targets (ATs are bigger than CHARA telescopes, and UTs even more). In the current paper, we focus on the
most promising science cases for development on the VLTI, taking into account the overall characteristics of the interferometric array at Cerro Paranal.


\subsection{Mass loss from evolved stars and galactic chemical enrichment}
\label{sec:SC2}

The chemical enrichment of our Galaxy is strongly influenced by Asymptotic Giant Branch (AGB) and Red Supergiant (RSG) stars, which supply heavy elements into the ISM through a mass-loss mechanism not yet fully understood.
This mass loss is certainly the result of a two-step process. First, the dust is formed behind shocks in an atmosphere extended by pulsation and/or convection. Then, radiation pressure on dust and/or molecules triggers the mass loss, in a complex mechanism that can involve magnetic fields.

Their surface is covered by a few convective cells \cite{chiavassa2010,montarges17}, evolving on time\-scales of several weeks to years \cite{chiavassa2011a}, and causing large temperature inhomogeneities. Their velocity can levitate the gas and contribute to the mass loss \cite{chiavassa2011b}.
Imaging of the surface of these stars is a key aspect to understand their dynamics and the link to the mass-loss process \cite{paladini2017}, with the possibility to resolve spatially and spectroscopically lines features in the photosphere.

The visible domain is still not fully explored at the high angular and spectral resolutions available with interferometers. Getting an interferometer equipped with a spectral resolution of about 30,000 on a wide bandwidth would enable studying large scale movements at the surface of the star, temperature stratification due to convection, and the kinematics of the cells.
In addition, the comprehensive interpretation of their surface dynamics, induced by pulsation and convection, requires multi-epochs observations.


With interferometry, one can determine the geometries of these stars, and the VLTI will clearly excel at it with its movable telescopes and imaging capabilities on $\approx10$\,mas targets. More and more observations indicate that giant stars and their descendant are not spherical. Determining the geometry of a substantial number of AGB stars is crucial to obtain accurate radiative transfer models and thus estimations of the mass-loss rate. This will will have to be a vital part of a community-large long-term multi-wavelength observing project aiming at determining the quantitative impact of AGB stars on the enrichment of the ISM.

Finally, a visible instrument will enable finding companions to AGB stars, and the VLTI is again very well suited for that with its 4 movable telescopes. Key binary parameters (mass and separation) can be extracted to quantitatively determine how binaries affect the mass loss from AGB stars and the Galaxy's chemical enrichment.

\subsection{A \& B stars binarity below 30\,au}
\label{sec:SC3}

De Rosa et al. \cite{VAST2014} have conducted a large volume-limited ($D < 75~pc$) multiplicity survey of A type stars, sensitive to companions beyond 30\,au, by using a combination of adaptive optics imaging and a multi-epoch common proper motion search. They reach the conclusion that $69.9\pm7.0\%$ of A-type stars have companions. Studies done on OB stars using AO and interferometry \cite{2014ApJS..215...15S} conclude to similar results.

Knowing the importance of hot stars for extragalactic stellar astrophysics (for example through early type eclipsing binaries in distant galaxies \cite{riess16,vilardell10}), and the fact that A-type stars have not yet been probed extensively with interferometry, we propose to go below this limit of 30\,au with the VLTI (down to 1\,au) to search for companions to A-type stars with the large baselines available at the VLTI array. This will permit us to correctly calibrate surface-brightness-color relationships of early-type stars, on which relies the distance determination of early-type eclipsing binaries \cite{challouf14}, and to remove potential biases due to the multiplicity of objects.

The main instrumental requirements for this program will be to reach magnitude up to $m_R=10$, long baselines (from 8 to 130\,m) and to be able to do fast snapshot observations with frequent revisits for the determination of the orbits. Therefore, the VLTI is uniquely suited to reach faint magnitudes and long baselines. Combining this survey with the Gaia data\cite{gaia}, we will be also able to determine an important number of stellar masses.

\subsection{Gaseous circumstellar environments : YSOs, Be and B[e]
stars, OBA winds, Cepheids, ...}
\label{sec:SC4}

Combining a spectrograph with an interferometer was one of the innovation of visible interferometry in the 70's \cite{labeyrie74}. This combination has proven to be extremely efficient at characterizing circumstellar environments where the gas is excited by a central ionizing source \cite{1989Natur.342..520M}, or masking underlying continuum emission \cite{ohnaka17}. Indeed, spectral lines analysis with an interferometer give access to the geometry, kinematics, chemical and physical conditions in the circumstellar medium at scales of a fraction of millisecond of arc.

In young stellar objects (YSOs), like Herbig or T Tauri stars, this corresponds to potential planet-forming regions. While optical direct imaging and sub-millimetric images revealed complex structures around YSOs (10\,au-scale spirals, gaps, holes), reaching the au-scale regions, where key star-disk-protoplanet(s) interactions processes occur, requires sub-mas angular resolution, only provided by visible interferometry.
Improving the sensitivity of visible interferometry and the ability to spatially resolve emission and absorption lines would be of strong interest, as one would gain in angular resolution and probe lines complementary to the infrared hydrogen Br$\gamma$ line. We can cite the hydrogen recombination lines (H$\alpha$, H$\beta$, ...) that are considered as accretion tracers, He\,I 1083\,nm that provides a unique diagnostic of kinematic motion in the regions close to the star, or forbidden lines (for instance [OI], [NII], [FeII], and [SII]) that trace low-density gas that is located on relatively large spatial scales.

In Cepheid stars, this corresponds to where lies the still-misunderstood circumstellar environments (CSEs). Indeed, a few of them observed by long-baseline interferometry show faint (few \%) and close (3 R$_{*}$) CSEs in the infrared.
Characterizing these CSEs, and in particular their radial brightness profile in several spectral bands will enable one to unbias the period-luminosity relation. Recently, $\delta$ Cep was observed with the VEGA/CHARA instrument and an unexpected visible CSE contributing to 7\% to the total flux was discovered \cite{nardetto16}. Many more visible CSEs can be discovered with a more sensitive visible instrument. Combining visible and infrared observations of Cepheids is probably one of the keys to characterize their environment.

In evolved emission-line stars, like Be stars or B[e] stars, this corresponds to where the disk is formed. This excretion disk is still a puzzle, as its exact formation process is not yet well understood. Non-radial pulsations, as well as the possible presence of a companion star, could be the culprits to trigger the ejection of matter around these stars. Visible interferometry can not-only fully characterize the gaz disk, but also detect unseen companion stars, and resolve the stellar disk itself.
Spectrally dispersed interferometry can be also applied to fast-rotating stars. The detection and characterisation of fast and/or non-solid rotation of these objects could provide decisive information on the inner structure of these stars, in which the internal redistribution of matter and evolution is strongly influenced by rotation-induced processes (e.g. gravity darkening, non-spherical mass loss, and internal circulation of mass, chemical elements, and angular momentum)

In O, B or A-type hot stars, this is where their radiative winds form. Emission lines seen in the visible can be resolved with the longest VLTI baselines, giving a direct access to previously-unconstrained physical parameters of the stellar wind. It is worth noting that, today, the VLTI is the only interferometer able to reach the southern sky and it exists many emission-line star studies in that region of the sky.

\subsection{Stellar masses and fundamental parameters of Main Sequence stars and Giants}
\label{sec:SC1}

A direct determination of the radius of the star is an important measurement complementing effective temperature and luminosity. These three properties are fundamental, and with only two independent, the third provides a method to test
systematic errors or to disentangle less-well-known variables that play a role in their determination e.g. line-of-sight absorption, distance, bolometric correction, surface gravity.
Having accurate determinations of these properties along with better constraints on their correlations results directly in a much more accurate and precise determination of the star's mass and evolution stage.

In particular, complementary information from, for example, asteroseismology, allows one to provide a direct estimate of the absolute stellar age, because the mass can be derived
with much higher accuracy.

Combining interferometry and asteroseismology has been possible for some stars using data for example from CoRoT, Kepler and K2 \cite{2004A&A...413..251K}\cite{2011A&A...534L...3B}\cite{2017MNRAS.471.2882W}. In a near future, TESS (NASA) and Plato (ESA) will provide seismic data on brighter sources and for a much larger number.

Focusing on the brighter stars allows effectively to determine an interferometric diameter. In addition, some of the stars whose diameters will be constrained will have an orbiting exoplanet detected by TESS. The constraints provided by interferometry will give the diameter and the age of the exoplanet.

On this particular science case, the VLTI will complete the SPICA survey (Northern hemisphere) with Southern stars, and especially those close to the galactic center.



\section{Instrumental requirements}
\label{sec:instru}

The high-level instrumental requirements that could be set with respect to the science goals presented above are the limiting magnitude to reach representative samples of stellar classes, the overall efficiency for rapid imaging and survey programs, the possibility to reach dense (u,v) coverage for imaging, and finally high to very high spectral resolution. Recent examples of stellar programs \cite{vedova,kluska} are driving our considerations.
The instrument is proposed as a dispersed fringes system with low (R=300) and medium (R=3000) spectral resolutions. A high (R=30000) spectral resolution Echelle spectrograph adapted to interferometric data is also considered. The instrument will take benefit of the installation of adaptive optics on all the telescopes and the light of each beam will be coupled into single mode fibers. This design is based on the experience gained on different instruments and on a prototype called FRIEND that we have installed on the CHARA Array (Martinod et al., 2018, in press).

Two main milestones are focusing our efforts. The first one is the optimization of the injection into single mode fibers at visible wavelengths with partial correction by adaptive optics. Coupling efficiency of at least 25\% should be reached and the stability of the coupling should be guarantied. Indeed combining 3, 4 or 6 beams for interferometric measurements suppose that the stability of the coupling on each beam is very high if one wants to reach a good interferometric efficiency. Photometric drops to 0 should be avoided, and optimized command-laws of the deformable mirrors should be implemented in this respect. 
The second milestone is the fringe stabilization to permit long exposure and thus to reach fainter magnitudes or lower fringe visibilities. The recent successes of the Gravity instrument \cite{gravity2017} are very encouraging. As a consequence, we study a fringe sensor similar to the Gravity one, adapted to shorter wavelengths.

\begin{figure}[htbp]
\centering
\includegraphics[width=0.9\textwidth]{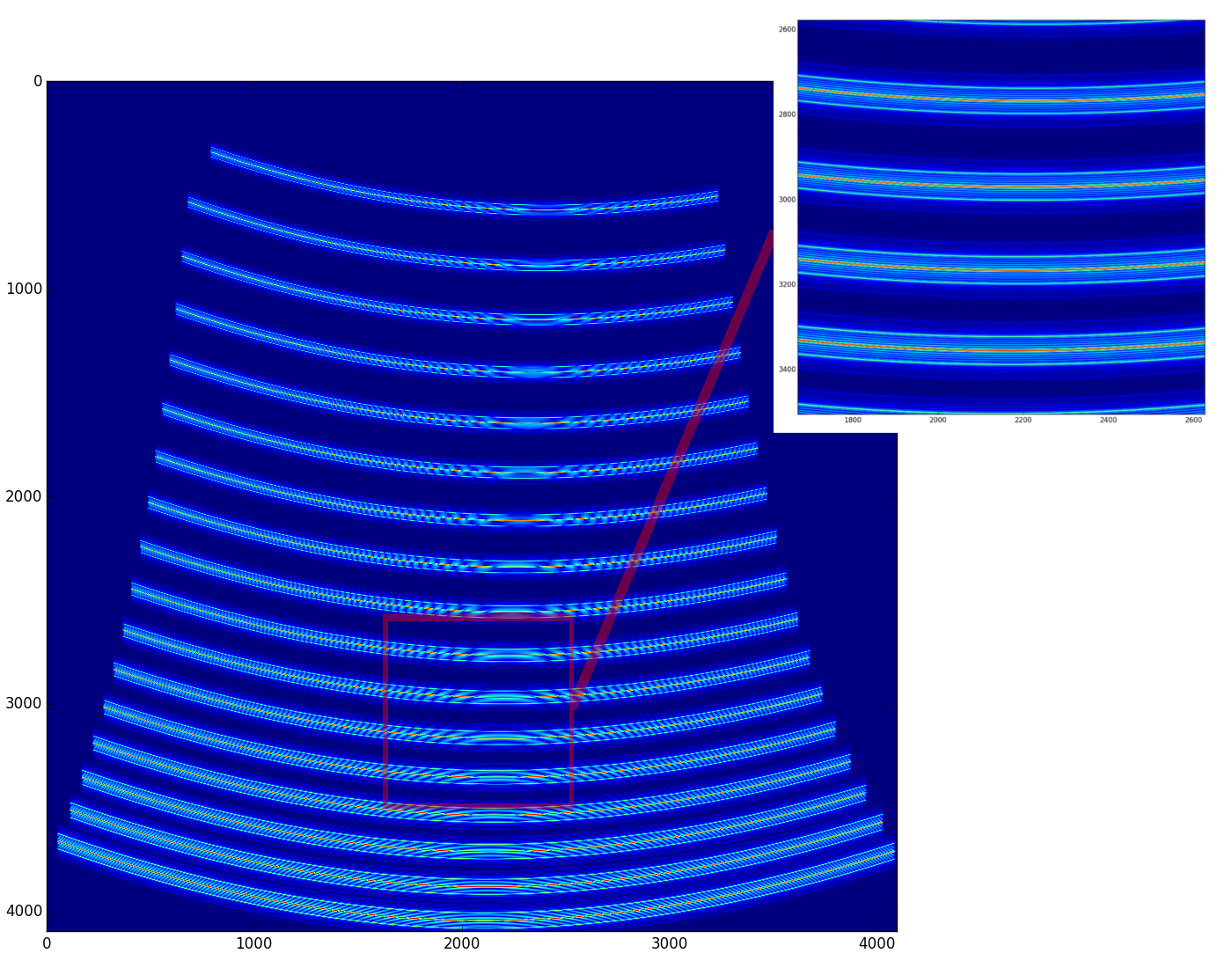}
\caption{Numerical simulation of a 4k detector fed with an Echelle spectrograph accommodating 17 orders of dispersion, each order being anamorphosed in order to correctly sample the high frequencies generated by the interferometric recombination of 6 pupils. }
\label{fig:spine}
\end{figure}

A high spectral resolution mode has been already used in the VEGA/CHARA instrument \cite{vega1,vega2} but with a limitation on the spectral bandwidth due to the sampling characteristics of this instrument. With a single-mode instrument in mind, we are considering an extension of the spectral band to a large fraction of the visible domain, by using the principle of Echelle spectroscopy coupled with the necessary anamorphic system required for the fringe sampling. The goal is to reach a spectral resolution R=$30000$, while keeping a band $\Delta\lambda=\SI{300}{\nano\meter}$ centered at \SI{700}{\nano\meter}. The resolution element is $\delta\lambda=\SI{0.02}{\nano\meter}$ and the sampling in the spectral direction should then be \SI{0.01}{\nano\meter}. A 6-telescope instrument generates 15 frequencies (in a linear non-redundant configuration) and each baseline will generate 4 fringes sampled over a minimum of $4\times3$ pixels. With all these numbers at hand, we can deduce that 17 orders of an Echelle spectrograph could fit in a 4K x 4K detector, as presented in Fig.~\ref{fig:spine}.

It is clear however that such a detector could not be read at the usual frequency in optical interferometry (of the order of the coherence time of the atmosphere) and thus long exposures of the order of 1s will be mandatory and are possible. This means that this mode should only be considered with the assistance of an active fringe tracker, which will put even more stringent requirements on the adaptive optics system.

The current VLTI is optimised for the infrared and therefore not ready to support a visible instrument yet. The adaptive optics beam splitters are dichroic mirrors optimised to only send the infrared light beyond \SI{1}{\micro\meter} to the VLTI instruments, setting aside visible light for the wavefront sensors of the adaptive optics systems. Permanently replacing the existing dichroic to send more visible light to the VLTI would affect the limiting magnitude of the adaptive optics and impact the red science carried out by the existing infrared instruments. The use of a reconfigurable dichroic in place of the current static one seems like the most promising solution. The installation of such a system in the existing star separators may turn out challenging, due to the crowded environment of the Coud\'e focus on both UTs and ATs.

The limited order of correction of the existing adaptive optics systems is an even greater challenge. Neither MACAO on the UTs nor NAOMI on the ATs have sub-apertures small enough to deliver a reasonable Strehl in the visible. Supporting our visible instrument with an infrared fringe tracker could make a low order adaptive optics system sufficient for fringe tracking, but the transmission to the visible instrument would remain dramatically low. On the ATs, the NAOMI adaptive optics system has the particularity of using an off-the-shelf mirror with an actuator count much larger than its wavefront sensor sub-aperture count. The actuator spacing of \SI{\sim16}{\centi\meter} is well adapted to the visible. However, in order to make full use of this deformable mirror, one would need to design a matching wavefront sensor. This could be done using the now decommissioned single-field Relay Optics Structure (ROS), the interchangeable interface between the telescopes and the tunnels to the delay lines. The dual-field ROS is the one currently in operation at VLTI; it contains the star separators and the NAOMI wavefront sensor. The nominal design of a visible compatible ROS would be based on a 11x11 wavefront sensor and require an upgrade to the real-time controller to support the correction order and bandwidth increase. It is worth mentioning that such a high order system on a small telescope in addition to supporting visible operation, would deliver an extremely high Strehl in the near-infrared, and especially at L band, in support to the high contrast Hi-5 endeavour\cite{Defrere+2018}. For the \SI{8}{\meter} UTs, the upgrade would be more extensive and also include the deformable mirror itself, since the MACAO deformable mirror, and its \SI{\sim1}{\meter} inter-actuator spacing, is not adapted to the visible.

\section{Study of the potential targets}
\label{sec:targets}

In order to study the number of potential targets accessible in the sky of Paranal ($-\SI{90}{\degree}< \delta < +\SI{20}{\degree}$), we made use of the SIMBAD database with different criteria in terms of angular diameters and magnitude in the R band.

A visible instrument addition will enable the VLTI to measure the angular diameter of several hundreds or thousands of stars. Indeed, an interferometer of \SI{150}{\meter} baseline at \SI{550}{\nano\meter} wavelength can fully resolve a star (i.e. attain the first zero of the Bessel function of a uniform disk) of \SI{0.9}{\milli\arcsecond} diameter. Stars smaller than this diameter can get their diameters measured, depending on the accuracy of the interferometer. We consider, based on the VEGA experience, that stars down to \SI{0.5}{\milli\arcsecond} can get their diameters measured. Crossing this information (and declination $\leq20^\circ $) with the JSDC catalog of predicted angular diameters (which contains itself 465,877 stars diameters), provides the number of $\approx 21000$ stars that can be resolved in the visible with the VLTI, given it can reach a limiting magnitude of $m_R<8$.

\begin{figure}[htbp]
\centering
\includegraphics[width=0.9\textwidth]{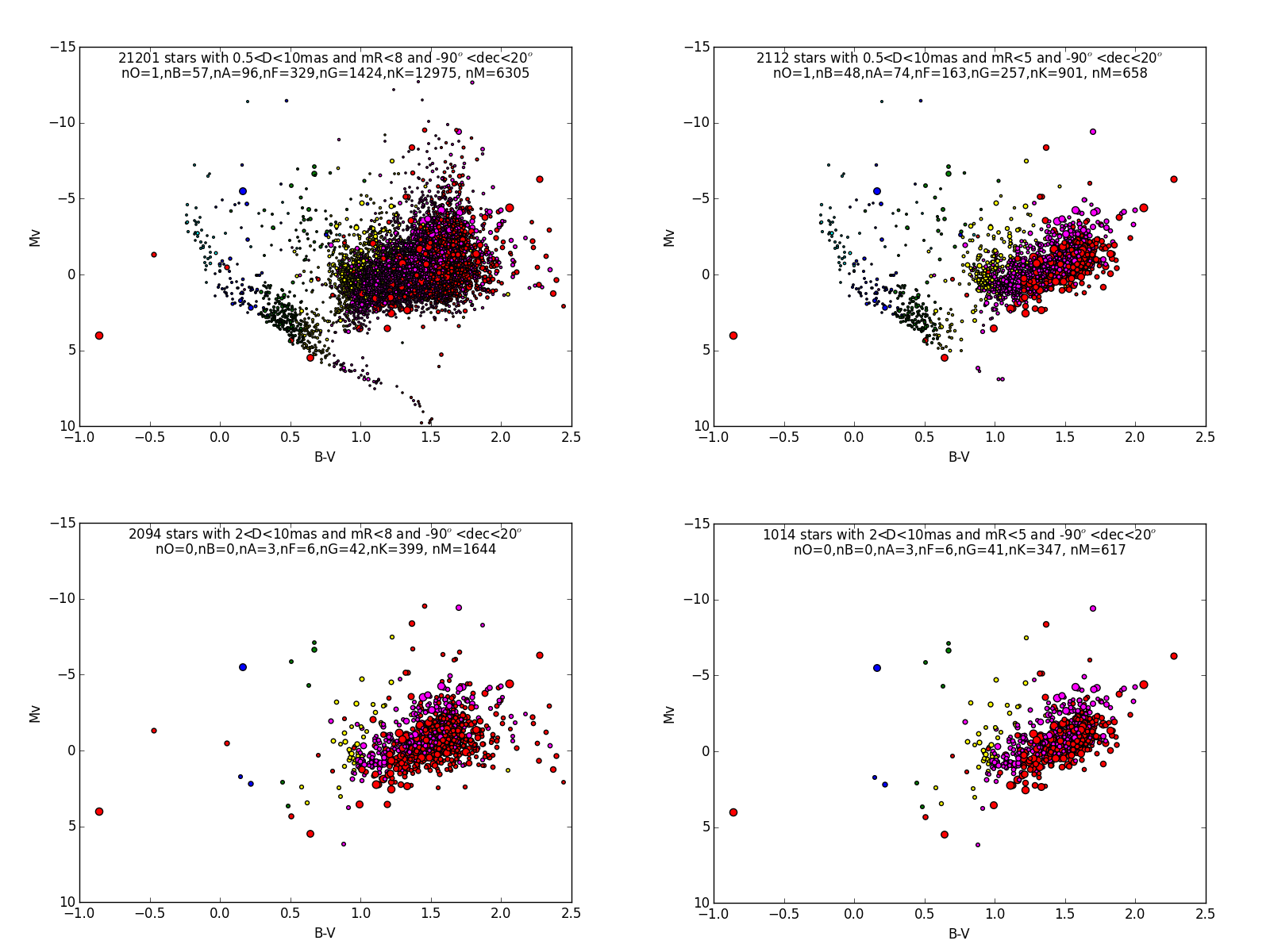}
\caption{{\bf Top row:} HR diagram for stars with $0.5<\theta(mas)<10$ and $m_R<8/5$ from left to right. {\bf Bottom row:} same but for $2<\theta<10$. The range of declination is $[-90^{\degree};+20^{\degree}]$.}
\label{fig:targets}
\end{figure}

However, this huge number has to be reduced depending on the sensitivity of the considered instrument. A $m_R<5$-limited instrument can resolve
$\approx~2000$ stars (i.e. almost all naked-eye visible stars), while a $m_R<3$-limited instrument can resolve $\approx 300$ stars.

The results are presented in Fig.~\ref{fig:targets} in the top line. The figure titles give numbers for each stellar type subset. This study shows that several hundreds of main sequence FGKM stars can get their diameters measured with a $m_R<5$-limited instrument on the VLTI already, with an over-representativity of K- and M-types stars. A $m_R<8$-limited instrument will outperform this by an order of magnitude, and the limit will rather come from available observing time than instrumental capacities.

On the other hand, imaging capabilities need the full angular resolution of a star. Therefore, the available sample is purposely limited to stars larger than \SI{2}{\milli\arcsecond} angular diameter (i.e. have at least 2 resolution elements on the star disk). As can be seen in Fig.~\ref{fig:targets} in the bottom line, this restricts the stellar types to mainly K and M stars, i.e. mostly AGB and RSG stars.

\begin{figure}[htbp]
\centering
\includegraphics[width=0.9\textwidth]{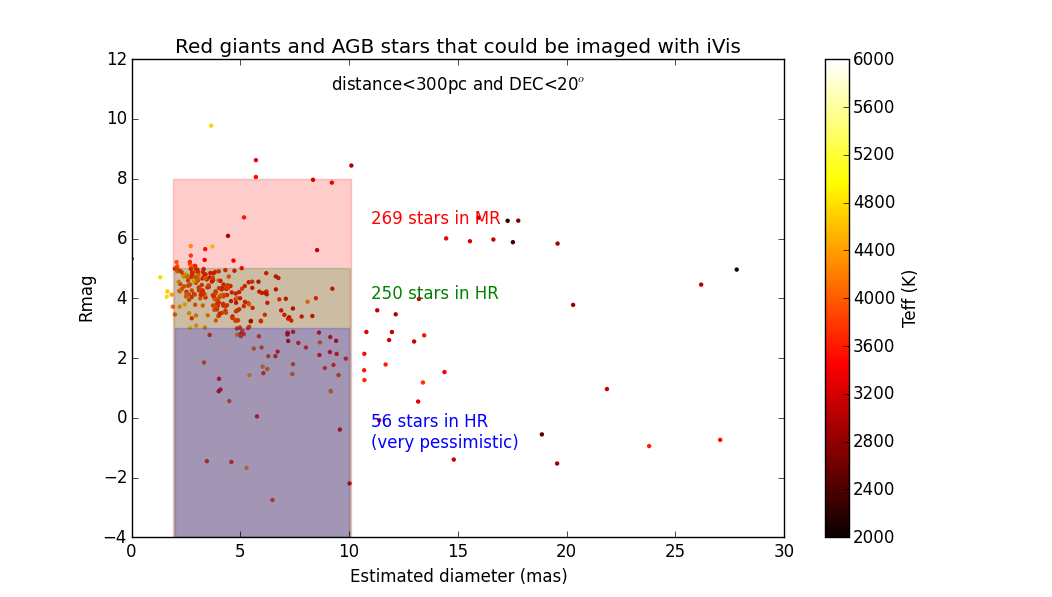}
\caption{Red giants and AGB stars that could be imaged with the visible instrument}
\label{fig:rgagb}
\end{figure}

To dig a bit further, we selected known RSG and AGB stars, and splitted them as a function of possible spectral resolution that can be used to observe them (the higher the spectral resolution, the worse the sensitivity for a given exposure time). The result of this study is presented in Fig.~\ref{fig:rgagb}. It shows that $\approx250$ RSG and AGB stars can be imaged with medium spectral resolution ($\approx1000$), and that a minimum of $\approx56$ RSG and AGB stars can be observed in high spectral resolution ($\approx30000$).

\section{Conclusions}
\label{sec:conclusion}
In this paper we have shortly presented our preliminary study of the potential for a visible interferometric instrument for the VLTI. Apart from additional considerations on the VLTI infrastructure for visible wavelengths, it is clear that the science goals presented above will require the assistance of a fringe sensor aiming at stabilizing the fringes in the visible. More detailed studies of the science programs will also be necessary in the future in order to correctly assess the need for the different spectral resolutions presented here. With our preparatory work on the CHARA Array, we continue to study critical aspects of this kind of instrument: optimization of the injection in the fibres, compensation of the birefringence, and preliminary study of a H-Band fringe sensor inspired from the Gravity one. We also plan to progress in the coming years on the Echelle spectrograph (SPINE) design in order to offer a wide spectral band at very high spectral resolution. This will clearly open new frontiers in stellar astrophysics.

\begin{acknowledgements}
This project has received funding from the European Research Council (ERC) under the European Union's Horizon 2020 research and innovation programme (grant agreement n$^\circ$730890). It was ade possible thanks to the OPTICON project, the Optical Infrared Coordination Network for Astronomy.

 We thank A. Domiciano de Souza for reading the paper and suggesting changes to the content.
\end{acknowledgements}

\bibliographystyle{spmpsci}      
\bibliography{spicavlti}   

\end{document}